\documentclass[twocolumn,prl,showpacs,superscriptaddress,preprintnumbers,amsmath,amssymb]{revtex4}

\usepackage[dvips]{graphicx}
\usepackage{dcolumn,color,epsfig}
\usepackage{bm}
\begin{document}

\title{Suppression of magnetism in Ba$_5$AlIr$_2$O$_{11}$: interplay of Hund's coupling,
molecular orbitals and spin-orbit interaction}

\author{Sergey V. Streltsov}
\affiliation{Institute of Metal Physics, S. Kovalevskoy St. 18, 620990 
Ekaterinburg Russia}
\affiliation{Ural Federal University, Mira St. 19, 620002 
Ekaterinburg, Russia}
\email{streltsov@imp.uran.ru}

\author{Gang Cao}
\affiliation{II. Physikalisches Institut, Universit$\ddot a$t zu K$\ddot o$ln,
Z$\ddot u$lpicher Stra$\ss$e 77, D-50937 K$\ddot o$ln, Germany}

\author{Daniel I. Khomskii}
\affiliation{II. Physikalisches Institut, Universit$\ddot a$t zu K$\ddot o$ln,
Z$\ddot u$lpicher Stra$\ss$e 77, D-50937 K$\ddot o$ln, Germany}

\begin{abstract}
The electronic and magnetic properties of Ba$_5$AlIr$_2$O$_{11}$ containing Ir-Ir dimers are investigated using the GGA and GGA+SOC
calculations. We found that strong suppression of the magnetic moment
in this compound recently found in [J. Terzic {\it et al.}, Phys. Rev. B {\bf 91}, 235147 
(2015)] is not due to charge-ordering, but is related to the joint effect of the spin-orbit 
interaction and strong covalency, resulting in the formation of metal-metal bonds.
They conspire and act against the intra-atomic Hund's rule exchange interaction
to reduce total magnetic moment of the dimer. We argue that the same mechanism 
could be relevant for other $4d$ and $5d$ dimerized transition metal compounds.
\end{abstract}

\maketitle

{\it Introduction.--}  The study of $4d$ and especially $5d$ transition metal compounds, in particular those of Ir, 
is now at the forefront of research in the physics of correated electron systems. This is 
largely connected with novel effects caused by strong spin-orbit coupling (SOC). In particular, 
for Ir$^{4+}$ with $t_{2g}^5$ electronic configuration, with spin $S=1/2$ and effective orbital 
moment $L_{eff}=1$, strong SOC can stabilise for an isolated ion the state with $J=1/2$, which can 
explain Mott insulating character of Sr$_2$IrO$_4$\cite{Kim2008} or could lead for honeycomb systems Li$_2$IrO$_3$ and Na$_2$IrO$_3$ to special states like those described by Kitaev model\cite{Jackeli2009} (see also Ref. \cite{Mazin2012}). 
But no less interesting  could be possible nontrivial properties of systems with Ir$^{5+}$ and Ru$^{4+}$, with 
ionic configuration $t_{2g}^4$ ($S=1$, $L_{eff}=1$), which in case of isolated ions are in 
nonmagnetic $J=0$ state\cite{Abragam}. And indeed for ESR (electron spin resonance) community 
Ir$^{5+}$ is a classical nonmagnetic ion, even sometimes used for nonmagnetic dilution. However, 
in concentrated solids the intersite interaction, if strong enough, can in principle lead to 
magnetic ordering in such systems -- the phenomenon known as singlet magnetism, see e.g. 
Ch. 5.5 in Ref. \cite{khomskii2014transition} and Ref. \cite{Khaliullin2013}. Apparently such magnetic state was discovered in double perovskite Sr$_2$YIrO$_6$ in Ref. \cite{Cao2014}.
%%Sergey: Ia ubral ssylki na Vas s Iogrem na Na2IrO3, t.k. oni ne ochen' podelu byli: oni vazhny dlia dannoi sistemy, bessporno, no vot ne v plane SOC, pro kotoruu my tut govorim.

An interesting system Ba$_5$AlIr$_2$O$_{11}$\cite{Lang1989} was recently experimentally studied in details in Ref. \cite{Terzic2015}. The main building blocks of it are dimers of face-sharing IrO$_6$ octahedra with, on the average, mixed valence Ir$^{4.5+}$, which may be expected to combine both the properties typical for Ir$^{4+}$ and for Ir$^{5+}$. However in contrast to the single-site physics (leading to $J=1/2$ state for Ir$^{4+}$ and $J=0$ for Ir$^{5+}$) here we deal with strongly coupled pair of Ir ions, in which, for example, intersite electron hopping can easily be of order or even larger than the intra-atomic parameters such as the Hund's rule coupling $J_H$ and spin-orbit coupling $\lambda$, and can compete with the intra-atomic Hubbard repulsion $U$. Indeed, in going from $3d$ to $4d$ and $5d$ ions, $U$ decreases, from $\sim$5 eV for $3d$ to $2-3$ eV for $4d$ and to $1-2$ eV for $5d$. Similarly, $J_H \sim 0.7-0.9$ eV for $3d$, $0.5-0.6$ for $4d$, and $\sim 0.5$ eV for $5d$ systems\cite{Sasioglu2011}. At the same time the size of $d-$orbitals, and with it the $pd-$ and $dd-$hoppings increase in this series, and can easily reach $1-1.5$ eV for $4d-5d$ systems\cite{Streltsov2012a,Kimber2013,Streltsov2015MISM}. In this situation there may occur strong modification of the behaviour expected for isolated $5d$ (e.g. Ir) ions. Ba$_5$AlIr$_2$O$_{11}$ may be a good example on which one can investigate relative importance of single-site vs intersite effects.
\begin{figure}[b!]
 \centering
 \includegraphics[clip=false,width=0.45\textwidth]{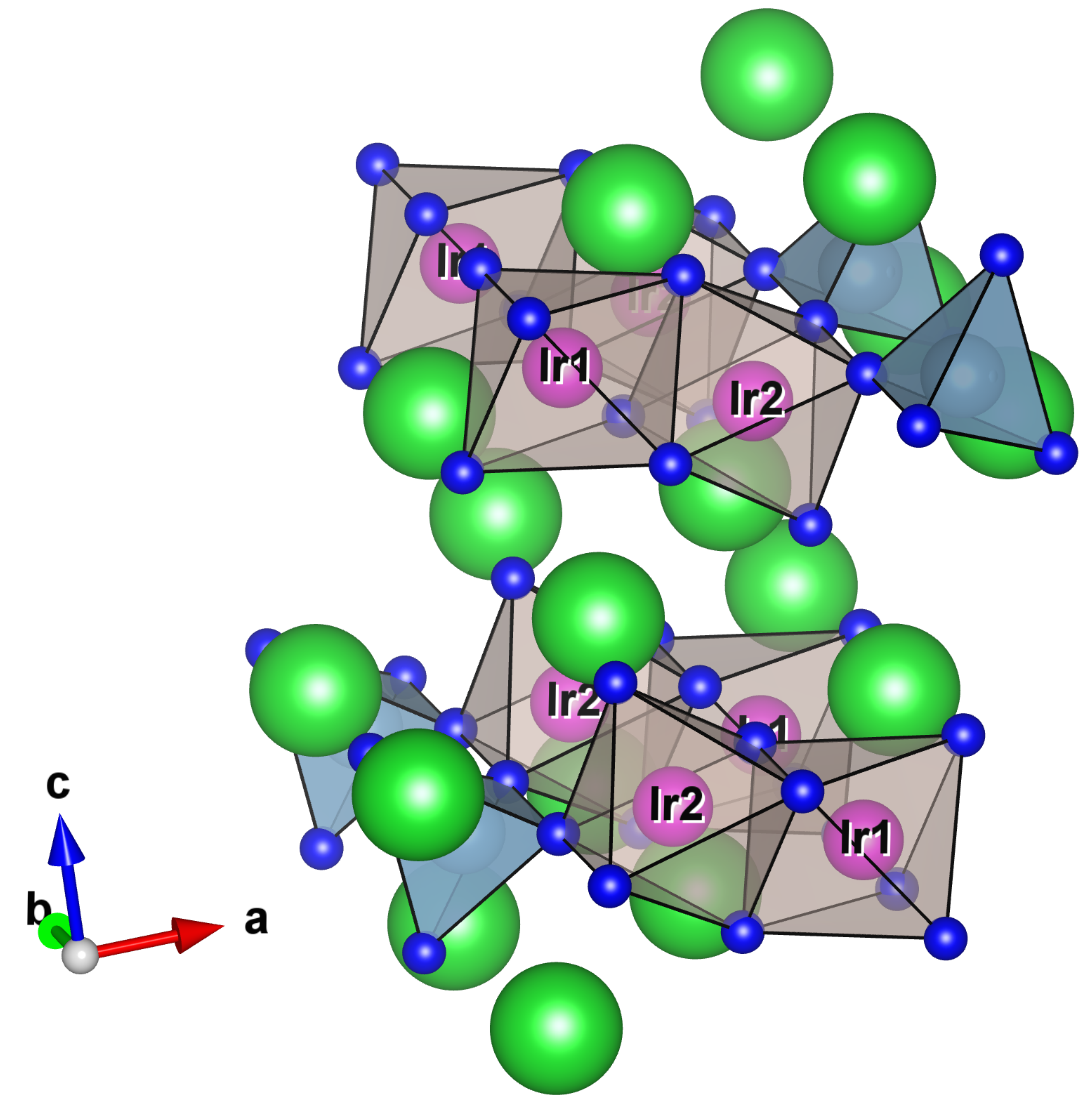}
\caption{\label{crystal-structure}(color online). Crystal structure of Ba$_5$AlIr$_2$O$_{11}$. Ir ions (violet balls) are in the oxygen (small blue balls) octahedra. Two nearest IrO$_6$ octahedra form
dimer, sharing their faces. Al (large blue balls) ions are in the oxygen tetrahedra and Ba (green balls) sits in the voids.}
\end{figure}

The crystal structure of Ba$_5$AlIr$_2$O$_{11}$ consists of
Ir-Ir dimers, which form chains,  as shown in Fig.~\ref{crystal-structure}.
At T$_S=210$ K there occurs a structural phase transition accompanied by the
metal-insulator transition\cite{Terzic2015}.  While even at room temperature there 
is a certain difference in the average Ir-O distance for two classes of Ir (Ir2 occupies 
octahedra, which share their corners with AlO$_4$ tetrahedra; Ir1 is in the center of the remaining octahedra), it increases at T$_S$. Thus, one could speak about certain charge 
ordering even for T$>$T$_S$, if this high temperature phase was insulating.
The real charge disproportionation in limiting case 2Ir$^{4.5+}\to$Ir$^{4+}+$Ir$^{5+}$, seems to occur only 
in the insulating phase below T$_S$ as manifested by a strong dielectric anomaly at T$_S$
and by increasing difference in the average Ir-O bond distance for two classes
of Ir.\cite{Terzic2015}

Below T$_M$=4.5 K there appears a long range magnetic order in Ba$_5$AlIr$_2$O$_{11}$
apparently an antiferromagnetic one, consistent with negative Curie-Weiss temperature ($\theta =-14$~K). The effective magnetic moment, obtained by the high temperature fit of susceptibility is $\mu_{eff}=1.04 \mu_B$/dimer,  much  smaller than one would expect from the values of spin moments corresponding
to Ir$^{4+}$ ($\mu_s = 1 \mu_B$/Ir) or Ir$^{5+}$ ($\mu_s = 2 \mu_B$/Ir) \cite{Terzic2015}. The mechanism of such a strong suppression was proposed in Ref. \cite{Terzic2015}. It was argued that is related to the joint effect of the strong spin-orbit coupling and formation of singlet molecular orbitals for part of Ir $5d$ orbitals. 

In this paper we theoretically investigate this problem using {\it ab initio} band structure calculations. We demonstrate that indeed in this material, as possibly also in other $5d$ compound, there exist strong interplay of covalent bond formation, Hund's rule coupling and spin-orbit interaction, which result in particular in strong suppression of magnetic moment of Ir ions and which strongly modifies intradimer exchange interaction.  These results give good explanation of unusual properties of Ba$_5$AlIr$_2$O$_{11}$, and show general trend expected in similar materials with competing intrasite and intersite effects.

\begin{figure}[t!]
 \centering
 \includegraphics[clip=false,width=0.5\textwidth]{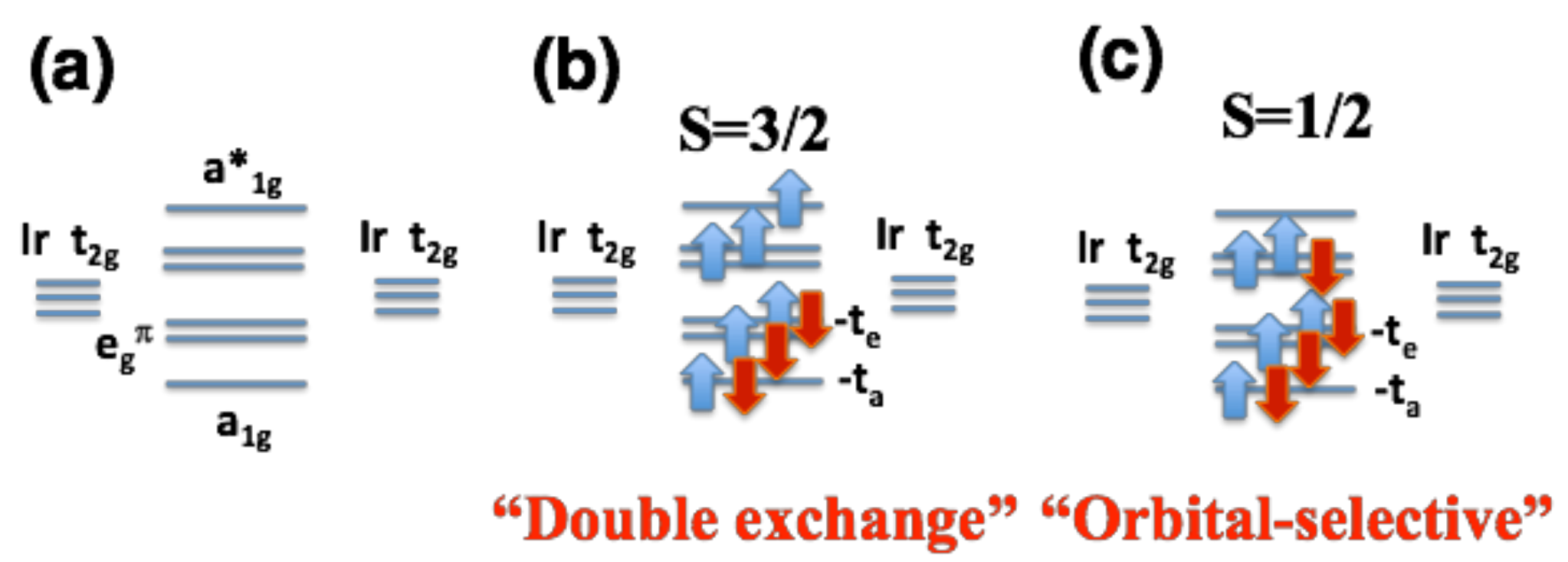}
\caption{\label{levels}(color online). The sketch, which shows (a) the level splitting in the dimer constructed out of face-sharing octahedra: the largest bonding-antibonding splitting corresponds 
to $a_{1g}$ orbitals, directed to each other in this geometry. (b) and (c) illustrates two possible
states in such a system with different values of total spin.}
\end{figure}

{\it Ionic treatment.--}
Before presenting the results of the real band structure calculations, we show what
one might expect  in this system starting from the ionic consideration. Since the $t_{2g}-e_g^{\sigma}$ crystal-filed splitting is huge for $4d$ and especially $5d$ transition metal oxides, first we have to fill $t_{2g}$ orbitals. Two neighbouring IrO$_6$ octahedra form dimer sharing their faces. In such geometry there will be two different by symmetry sets of orbitals: $a_{1g}$ orbitals pointing to each other will have stronger hopping, $t_{a}$, than $e_g^{\pi}$ orbitals, $t_{e}$, see Fig.~\ref{levels}(a)\cite{Kugel2015}. Having nine $5d-$electrons per Ir-Ir dimer one may fill these orbitals in two different ways: to have maximum ($S_{tot}=3/2$) and minimum ($S_{tot}=1/2$) total spins, Fig.~\ref{levels}(b) and (c) respectively.

The first configuration with $S_{tot}=3/2$ can be called double exchange (DE) state, since the electron (hole) on delocalized $a_{1g}$ antibonding orbital with the largest hopping $t_a$ moves from one site to another in the dimer and makes other two electrons (holes) to have the same spin projection. In the second, state with $S_{tot}=1/2$ the antibonding $a_{1g}$ orbital stays unoccupied and the total magnetic moment is suppressed. One may speak about this state as orbital-selective (OS) states 
\cite{Streltsov2014,Streltsov2015MISM}, since $e_g^{\pi}$ and $a_{1g}$ orbitals behave very differently in this state.

One may consider this situation in the simplest ionic approximation, taking into account Hund's rule coupling $H_{Hund} = - J_H (1/2 + 2 \vec S_1 \vec S_2)$ and the hoppings between $a_{1g}$ orbitals ($t_a$)  and $e_g^{\pi}$ orbitals ($t_e$). We also assume in the beginning that the wave-function for $a_{1g}$ electrons in the OS state can be approximated by pure molecular-orbital (MO LCAO) state (homopolar and ionic terms have the same weights).
Then it is easy to show that the DE state will be realized, if
\begin{eqnarray}
\label{ineq}
J_H > 2 (t_a - t_e) = 2\Delta_{ae}.
\end{eqnarray}
The Hund's rule exchange for Ir is $\sim$0.5-0.7 eV\cite{Foyevtsova2013,VanderMarel1988}, while both hopping parameters can be found from real {\it ab initio} calculation. If $\Delta_{ae}$ would be
large enough, one could explain experimentally observed suppression of magnetic moment
only by the covalency, i.e. by formation of metal-metal bonds. 

{\it Calculation details.--}
We used full-potential Wien2k code\cite{Blaha2001} and generalized gradient approximation (GGA)\cite{Perdew1996}.  The atomic sphere radii were set as following: R$_{Ir}$=1.91 a.u., R$_{Ba}$=2.35 a.u., R$_{Al}$=1.63 a.u., and R$_O$=1.63 a.u. The spin-orbit coupling (SOC) was treated in a second variational way. 
160 k-points were used for the Brillouin-zone integration. The parameter of the plane-wave expansion was chosen to be R$_{MT}$K$_{max}$ = 7, where R$_{MT}$ is the smallest atomic sphere radii and K$_{max}$ - plane wave cut-off. The calculations were performed for the crystal structure obtained by X-ray diffraction at T=90 K \cite{Terzic2015,SGK-sup}.

\begin{figure}[t!]
 \centering
 \includegraphics[clip=false,width=0.5\textwidth]{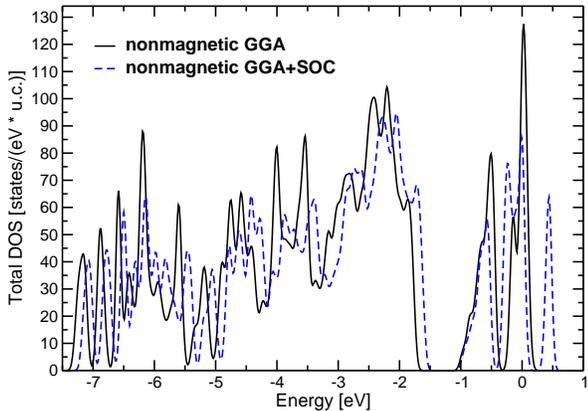}
\caption{\label{GGA_nm}(color online). Total DOS in the nonmagnetic GGA and GGA+SOC calculations. In the GGA calculations bonding and antibonding Ir orbitals are at -0.7 and -0.1 eV, respectively. }
\end{figure}

{\it Calculation results.--}
Our nonmagnetic GGA calculation for low-temperature phase indeed
indicates a sizeable bonding-antibonding splitting (see Fig. \ref{GGA_nm}), which is natural for IrO$_6$ octahedra forming dimers. As we have seen above, the key parameter, which defines the ground state
electronic configuration is the splitting between antibonding $a_{1g}$ and $e_g^{\pi}$ orbitals, $\Delta_{ae}$.
Using the linearized muffin-tin orbitals method\cite{Andersen1984}, the local density approximation and Wannier projection technique\cite{Streltsov2005} we estimated, that $\Delta_{ae} \sim $0.2 eV.  In contrast to our expectations, this value is smaller than $J_H/2$. Therefore in contrast to experimental finding \cite{Terzic2015} according to Eq. \eqref{ineq} the DE, not OS state with small magnetic moment  should win in this case.

Indeed, in the magnetic GGA calculations total spin moment is $\sim$2.0 $\mu_B$/dimer (smaller than the ionic value due to hybridization effects\cite{Streltsov2013a,Cao2002,Streltsov2008a}), while $|\mu_s$(Ir1)$|$=0.9 $\mu_B$ and $|\mu_S$(Ir2)$|$=0.6 $\mu_B$. It is remarkable that the spin moments on the two Ir ions forming dimers are ferromagnetically coupled (antiferromagnetic solution does not converge in the GGA). Therefore, the exchange coupling between these ions without SOC is governed by the DE. Furthermore, there is an unusually large moment $\sim$0.5 $\mu_B$/dimer in the interstitial space between the atomic spheres related to the formation of the bonding state, favoring bond-centered spin densities.

However, there is a significant difference between the two unequivalent mean Ir-O distances for the two face-sharing octahedra: 
$\delta$=$d$[Ir2-O]-$d$[Ir1-O]=0.016 \AA ~(recalculated from experimental structure in low-temperature phase\cite{Terzic2015,SGK-sup}), compared to $\delta \sim$ 0.055 \AA~for a full Ir$^{4+}$/Ir$^{5+}$ charge order\cite{Shannon-76}. The Bader analysis \cite{bader1994} shows that corresponding charge disproportionation is $\delta n_{Ir1/Ir2} \sim 0.3$ electrons (Ir1 is closer to Ir$^{5+}$ and Ir2 to Ir$^{4+}$), indicating the existence of a charge order.

\begin{figure}[t!]
 \centering
 \includegraphics[clip=false,angle=270,width=0.5\textwidth]{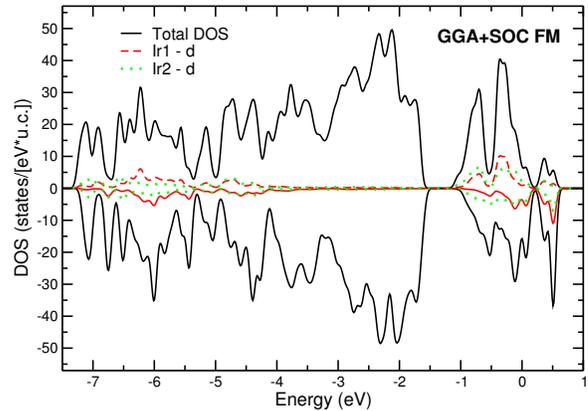}
\caption{\label{GGASOC}(color online). Results of the ferromagnetic GGA+SOC calculations. Positive (negative) values correspond to spin majority (minority). E$_F$=0. }
\end{figure}

An account of the SOC in the magnetic GGA+SOC calculations strongly changes the situation. It reduces the total moment, $\mu_z^{GGA+SOC} \sim 0.8 \mu_B$/dimer, which is much smaller than in GGA, where $\mu_z^{GGA} \sim 2 \mu_B$/dimer, and which is now consistent with the experimental value. This suggests the importance of the SOC. However, the SOC does not simply reduce the total moment due to direct contribution of orbital moment, which is expected to be antiparallel to spin, see Tab. \ref{moments}. This effect, commonly used for the description of the spin singlet state of Ir$^{5+}$ ion (which for isolated ion could give nonmagnetic state \cite{Abragam,Khaliullin2013,khomskii2014transition}),
 leads in Ba$_5$AlIr$_2$O$_{11}$ to a decrease of the total moment only by $\sim$0.2 $\mu_B$/dimer. Thus the observed reduction of the total moment of a dimer is not caused by onset of $J=0$ state on Ir$^{5+}$. This is due to the fact that we are dealing not with the isolated ions, but with a dimer, with significant hopping between sites and with the average mixed valence of Ir$^{4.5+}$. It is clearly seen from Fig.~\ref{GGASOC}, that $5d$ orbitals of Ir1 and Ir2 are strongly hybridized and cannot be considered as ionic. The main reason for the reduction of the total moment is related to strong changes in the electronic structure and to breaking of the delicate balance between DE and SO states by the SOC.

These changes are easier to see in the nonmagnetic GGA+SOC calculations. One may notice in Fig.~\ref{GGA_nm} that the SOC basically shifts part of the antibonding MO to higher energy, due to formation of $j_{eff}=1/2$ and $j_{eff}=3/2$ subbands. The DOS center of gravity calculations shows that the splitting due to SOC is $\Delta_{SOC} \sim 0.6$ eV. This, together with the bonding-antibonding splitting  is already sufficient to overcome the Hund's rule coupling and to suppress DE. Indeed, it is clearly seen in Fig.~\ref{GGASOC} that the SOC does not spoil main feature of the GGA band structure -- the presence of bonding-antibonding splitting, but additionally lifts one of the antibonding orbitals up 
so that in effect $\Delta_{ae}+\Delta_{SOC} >J_H/2$, cf. Eq~\eqref{ineq}. Thus, the SOC plays on the side of covalency against DE. It also decreases the moment in the interstitial region down to 0.27 $\mu_B$ and mixes spin up and down states reducing spin moments on Ir sites, as shown in Tab. \ref{moments}. On the other hand the SOC does not act against charge disproportionation, which is given by the lattice distortions: $\delta n_{Ir1/Ir2}$ stays $\sim$0.3 electrons in the GGA+SOC calculations.

	These theoretical results are consistent with experiment. Particularly,  considerable weakened 
$\mu_{eff}$  is a result of common action
of the SOC and covalency. As mentioned above, Ir forming dimer should not be considered as isolated ions, but they rather represent a single quantum-mechanical object having, due to joint effect of the SOC and covalency, strongly reduced magnetic moments. These moments can be coupled between dimers antiferromagnetically as usually occurs in insulating TM oxides\cite{khomskii2014transition}. This agrees with the low temperature of the magnetic transition and negative $\theta_{CW}$. It is important to note that small moment is experimentally seen already at temperatures much higher than the temperature of charge ordering; therefore it is not related with the formation of Ir$^{4+}$ and Ir$^{5+}$ ions (and correspondingly with $J=0$ physics), but 
is rather explained by the competition between 
(covalency +SOC) with the Hund's exchange. 
\begin{table}
\centering \caption{\label{moments} Magnetic moments obtained in the GGA+SOC calculations.}
\vspace{0.2cm}
\begin{tabular}{lcccc}
\hline
\hline
Ion     & Spin moment, $\mu_B$ & Orbital moment, $\mu_B$  & $\mu_j$(Ir), $\mu_B$\\
\hline
Ir1 (Ir$^{5+}$) & (0.02, 0.00, 0.53) & (-0.08, 0.00, -0.09) & 0.44\\
Ir2 (Ir$^{4+}$) & (0.09, 0.00, 0.24) & (-0.01, 0.00, -0.11) & 0.15\\
\hline
\hline
\end{tabular}
\end{table}

{\it Conclusions.--} To sum up, with the use of the {\it ab initio} calculations
we show  in the present paper that it is the combined action of the spin orbit coupling and strong covalency which leads to suppression of magnetic moment in Ba$_5$AlIr$_2$O$_{11}$. Formation of the metal-metal bonds (covalency) alone is here not strong enough to suppress double exchange, which would favours the state
with maximum spin. The spin-orbit coupling alone also would not be able to efficiently suppress
magnetic moments on Ir, due to not complete Ir$^{4+}$/Ir$^{5+}$ charge-ordering
apparently caused by strong intersite electron hopping. Only combined action of both these mechanisms leads to the state with properties observed experimentally. We suppose that similar situation may also be met in other transition metal compounds, especially those with $5d$ ions.

{\it Acknowledgments.--}
SS is grateful to Dr. Kateryna Foyevtsova for useful discussions. This work was supported by National Science Foundation via Grant No. DMR-1265162 (GC), the U.S. Civilian Research and Development Foundation (CRDF Global) via Grant FSCX-14-61025-0 (SVS, GC), Russian Foundation of the Basic Research via Program 16-32-60070 (SVS), FASO (theme``electron'' 01201463326) and MON (project 236), Cologne University via German Excellence Initiative and German project FOR 1346 (DK).

\bibliography{../library}
\end{document}